\documentclass[12pt,preprint]{aastex}

\def\HII{{\ion{H}{2}}}
\def\mum{$\mu$m}

\newcommand{\Msunpyr}{\ifmmode {\rm\,M_\odot\,yr^{-1}} \else {${\rm\,M_\odot\,yr^{-1}}$}\fi}
\newcommand{\OmM}{\ifmmode {\Omega_{\rm M}}\else $\Omega_{\rm M}$\fi}
\newcommand{\OmL}{\ifmmode {\Omega_{\Lambda}}\else $\Omega_{\Lambda}$\fi}
\newcommand{\kmps}{\ifmmode {\rm\,km~s^{-1}} \else ${\rm\,km\,s^{-1}}$\fi}

\received{}
\begin{document}

\title{Modelling the Pan-Spectral Energy Distribution of Starburst Galaxies: II.\\
Control of the \HII\ Region Parameters}

\author{Michael A. Dopita,  J\"org Fischera, \& Ralph S. Sutherland}
\affil{Research School of Astronomy \& Astrophysics,
The Australian National University, \\
Cotter Road, Weston Creek, ACT 2611, Australia}
\author{Lisa J. Kewley,}
\affil{University of Hawaii at Manoa, Institute for Astronomy, 2680 Woodlawn Drive, Honolulu, HI 96822, USA}
\author{Richard J. Tuffs, Cristina C. Popescu, }
\affil{Max-Plank-Institut f\"ur Kernphysik, Saupfercheckweg 1, D-69117 Heidelberg, Germany \&
Observatories of the Carnegie Institution of Washington, 813 Santa Barbara St., Pasadena, CA 91101, USA}
\author{Wil van Breugel}
\affil{Institute of Geophysics and Planetary Physics, Lawrence Livermore National Laboratory, L-413
Livermore, CA 94550, USA}
\author{Brent A. Groves}
\affil{Max-Plank Instute for Astrphysics, Karl-Schwartzschild-Str 1, 85741, Garching, Germany}
\author{\& Claus Leitherer}
\affil{Space Telescope  Science Institute, 3700 San Martin Drive, Baltimore  MD21218, USA}
\email{Michael.Dopita@anu.edu.au}


\begin{abstract}
We examine, from a theoretical viewpoint, how the physical parameters of \HII\ regions are controlled in both normal galaxies and in starburst environments. These parameters are the \HII\ region luminosity function, the time-dependent size, the covering fraction of molecular clouds, the pressure in the ionized gas and the ionization parameter. The factors which control them are the initial mass function of the exciting stars, the cluster mass function, the metallicity and the mean pressure in the surrounding interstellar medium. We investigate the sensitivity of the H$\alpha$ luminosity to the IMF, and find that this can translate to more than a factor two variation in derived star formation rates. The molecular cloud dissipation timescale is estimated from a case study of M17 to be $\sim1$~Myr for this object. Based upon \HII\ luminosity function fitting for nearby galaxies, we suggest that the \HII\ region  cluster mass function is fitted by a log-normal form peaking at $\sim 100 M_{\odot}$. The cluster mass function continues the stellar IMF to higher mass. The pressure in the \HII\ regions is controlled by the mechanical luminosity flux from the central cluster. Since this is closely related to the ionizing photon flux, we show that the ionization parameter is not a free variable, and that the diffuse ionized medium may be composed of many large, faint and old \HII\ regions. Finally, we derive theoretical probability distributions for the ionization parameter as a function of metallicity and compare these to those derived for SDSS galaxies.
\end{abstract}


\section{\label{intro}Introduction}
In distant galaxies and starbursts, much of what we infer about the star formation rates, the physical conditions in their interstellar medium (ISM), and the chemical abundances (or ``metallicity") in the ISM is determined by a study of their emission line spectra. However, we must be constantly aware that what we observe are the integrated properties of the population of \HII\ regions in these galaxies, and the integration extends over a wide range of \HII\ region ages, cluster masses, sizes, pressures, ionization parameters and potentially, over a wide range of metallicity as well. 

In nearby disk galaxies the individual \HII\ regions can be spatially resolved and subjected to independent study, but the properties of these \HII\ regions are still intimately connected to the properties of their local ISM and to the nature and the state of evolution of their central cluster. 

In both cases, it would be wise for us to attempt to understand how the physical parameters of the \HII\ regions and their emission line spectra are controlled by their cluster, their environment, and their metallicity. Surprisingly, there has been very little effort devoted to this problem. In general, the modelling of the emission line spectrum of distant galaxies has been treated as if it originated from a single \HII\ region of a given metallicity and ionization parameter. This ionization parameter is defined as either the ratio of the mean photon flux to the mean atom density, $q$ or in its dimensionless form as the ratio of mean photon density to mean atom density ${\cal{U}}=q/c$. The excitation of the \HII\ region is very sensitive to both the IMF and the age distribution of the exciting stars. In the modelling hitherto, this is usually treated in two limits; either the instantaneous burst approximation or in the continuous star formation approximation in which stars are supposed to be born continually within the ionized region (see \citet{Kewley01a}, and references therein).

Clearly, this modeling approach is rather crude, and does not provide insight into the dynamical and age evolution of the individual clusters and their associated \HII\ regions. These are important parameters because they determine the geometry of the gas with respect to the exciting stars. The geometry determines both the excitation of the \HII\ region and the radiation field density incident on photodissociation region (PDR) around the \HII\ region. This in its turn determines the characteristic dust temperature, and hence the wavelength of the peak in the FIR dust re-emission feature. 

As discussed in \citet{Dopita05}, hereinafter referred to as Paper I, for starburst galaxies dominated by their young stellar population and possessing an ISM of high density, the global SED can be adequately be described by the sum of the SEDs of its embedded \HII\ regions. This approximation requires that the radiative transfer problem can be approximated as a local, rather than as a global problem. Such a model is correct provided that the majority of the far-IR emission arises from absorption of the UV radiation field in a relatively thin dust layer identified with the classical photo-dissociation region (PDR). These typically have an optical depth corresponding to $A_{\rm V} \sim 3$, and a thickness $dR \sim 300/n_{\rm H}$ pc. In the molecular regions surrounding normal galactic  \HII\ regions, densities are typically $100 - 1000$ cm$^{-3}$, implying that much of the far-IR which they produce comes from a layer of parsec or sub-parsec dimensions. In starburst galaxies, interstellar pressures may range up to a factor of 100 higher than in normal disk galaxies, and these produce correspondingly thinner PDR zones. The assumption that the PDR is thin justify the use of models which treat the radiative transfer as either localized within or immediately surrounding the \HII\ regions.

A theory for the size distribution of \HII\ regions has been presented by \citet{Oey97} and \citet{Oey98}.
According to these papers, the \HII\ regions expand and evolve as mass-loss and supernova energy driven bubbles for as long as their internal pressure exceeds the ambient pressure in the ISM. When this condition is no longer met, the \HII\ region is assumed to ``stall".  The time taken to reach the stall radius turns out to be proportional to the stall radius, and the mass of the OB star cluster is key in determining the stall radius; low mass cluster \HII\ regions stall early and at small radius. The resultant size distribution therefore depends strongly upon the cluster mass distribution, which can in principle de derived from the observed \HII\ region H$\alpha$ luminosity function. In the \citet{Oey98} paper, this luminosity function was modelled on the assumption that the number of ionizing stars was a power-law, $N(N_*)dN_*=BN_*^{-\beta}dN_*$. In a later paper, \citet{Oey03} suggest that the differential nebular size distribution can be described by a power law of slope -4, with a flattening for sizes below $\sim130$~pc, in contrast to the exponential size distribution usually adopted.

During the expansion phase the \HII\ region is treated as a bubble powered by the mechanical energy input of the mass-losing central stars, and following the classical \citet{Castor75} theory.The radius $R$ and internal pressure,  $P$, of a mass-loss and/or supernova explosion pressurised \HII\ region are then given by  \cite{Oey97, Oey98};
\begin{equation}
R = \left(\frac {250}{ 308\pi} \right)^{1/5}L_{\rm mech}^{1/5}\rho_0^{-1/5}t^{3/5} \label{R}
\end{equation}
and
\begin{equation}
P = {\frac{7}{(3850\pi)^{2/5}}}L_{\rm mech}^{2/5}\rho_0^{3/5}t^{-4/5}. \label{P}
\end{equation}
Here, $L_{\rm mech}$ is the instantaneous mechanical luminosity of from the central stars from all sources, $\rho_0$ is the density  of the ambient medium and $t$ is the time. The particle density is given in terms of the ambient density by $n_0=\rho_0/\mu m_H$, and the ambient pressure $P_0=nkT_0$. Eliminating $t$ between equations \ref{R} and \ref{P}, we obtain the pressure in terms of the radius of the \HII\ region, assuming that the ionized plasma layer is thin compared with the total radius:
\begin{equation}
P = {\frac {7 } {(3850\pi)^{2/5}}}\left(\frac{250}{ 308\pi }\right)^{4/15}{\left(\frac {L_{\rm mech}}{ \mu m_Hn}\right)}^{2/3} {\frac{\mu m_H n_0}{ R^{4/3}}} \label{P1} 
\end{equation}

This theory emphasizes the important role of the pressure in the ISM in determining the radius of
an individual \HII\ region, and its evolution with time. In particular, with higher ISM pressure, the 
\HII\ region stalls at smaller radius, and therefore the mean dust temperature in the atomic and molecular
shell around the \HII\ region should be higher in high-pressure environments. This relationship was investigated in detail in  Paper I, but the models presented there were for only a single cluster mass and single set of chemical abundances. However, the dynamical evolution evolution was treated in a more sophisticated self-consistent manner which allowed for the time-dependence of both the mass-loss driven mechanical energy input of the central stars, and the energy input from supernova explosions.

Equation \ref{R} shows that the size of an \HII\ region is correlated, albeit weakly, with the $L_{\rm mech}$ of the cluster, that is to say, with cluster mass, metallicity and with the age of the cluster. The radius of the \HII\ region and the properties of the exciting cluster together determine the characteristic peak of the far-IR dust re-emission bump. Likewise, equation \ref{P1} shows that the ionization parameter of the \HII\ region is determined not only by the flux of ionizing photons from the cluster (which is a function of metallicity, cluster mass and time; $N_*(Z, M_{clust}, t)$ but also by the radius of the \HII\ region and by $L_{\rm mech}$. 

In this paper, we use \emph{Starburst 99 v.5} \citep{Leitherer99} stellar synthesis models to investigate the effect of metallicity, the stellar initial mass function and the cluster mass function in controlling the excitation of the populations of \HII\ regions in both disk and starburst galaxies galaxies. In particular, we will investigate what constraints are placed by observations of H$\alpha$ luminosity functions upon the cluster mass functions, and what are the probability distributions of \HII\ region ionization parameters in galaxies as a function of metallicity and pressure in the ISM.

In the third paper in this series, we will investigate what effect these have upon the emission line diagnostics of ensembles of \HII\ regions in galaxies with the view to assisting the analysis of the strong line emission spectra of distant unresolved galaxies in order to derive star formation rates, metallicities and ISM pressures.

\section{The Stellar IMF and Star Formation Rates}
It has often been remarked that the form of the stellar initial mass function (IMF) presents severe observational obstacles in the path of observers who seek to determine it. The slope of the initial mass function, defined as the frequency distribution of stellar masses per unit logarithmic interval of mass;  $\Gamma = dN(logM)/dlogM$ is a steeply decreasing function of mass, of order -2.35 for the \citet{Salpeter55} IMF. This ensures that most of the stellar mass is concentrated into low-mass stars. As for the luminosity, the steep stellar mass-luminosity relation more than compensates the downward-weighting effect of the IMF. Stars capable of ionizing the surrounding ISM follow mass-luminosity relation;  $L \propto M^{1.5-2.0}$  \citep{Schaller1992}, with the smaller exponent applicable to higher masses. The combined effect of the IMF and the mass-luminosity relation makes O stars with masses $\sim 50 M_{\odot}$ the typical providers of ionizing photons in a standard star cluster.

This affects all attempts to calibrate the rate of star formation (M$_{\odot}$~yr$^{-1}$), using the observed flux in a hydrogen recombination line.  The H$\alpha$ line is usually used for this purpose, e.g. \citet{Dopita94,Kennicutt98,Kewley02,Panuzzo03} and Paper 1.  The calibration depends sensitively on the mass ratio of the massive stars which produce ionizing photons to the total mass of the stars being born, and  is determined by the shape of the IMF. Estimates of the star formation rate based upon the H$\alpha$ line are also affected by dust absorption in the regions surrounding the \HII\ regions, or by dust absorption within the \HII\ region where dust competes successfully with dust to absorb the ionizing photons \citep{Petrosian78, Panagia74, Mezger74, Natta76, Sarazin77, Smith78, Shields95, Bottorff98,Inoue00, Inoue01a, Inoue01b, Dopita03}.

Similarly, the calibration of the rate of rate of star formation using the the total 8-1000~$\mu $m infrared luminosity, $L_{{\rm IR}},$ as measured in the rest frame frequency of the galaxy can also strongly affected by the form assumed for the IMF. This will show a different dependence on IMF than for estimates of star formation based upon the the H$\alpha$ line because stars providing the visible and UV luminosity which is dust-absorbed are of lower mass, on average, than those which produce the ionizing photons.

In its current 2005 version, the \emph{Starburst 99} code allows us to investigate what effect different assumed IMFs will have upon these calibrations. The code allows the IMFs to be approximated as a series of piecewise power-laws. We have therefore investigated (over the mass range $0.1- 120M_{\odot}$) three commonly-used forms of the IMF, the original \citet{Salpeter55}, the \citet{M-S79} form, and the \citet{Kroupa02} formulation. We have used piece-wise power-law fits to represent these IMFs, although we note that their paper, \citet{M-S79}  represented their IMF as a truncated lognormal distribution (with its maximum at zero). However, they also represented it as a three-segment broken power law, which we have used here because it is the form of the IMF which Starburst 99 currently accepts. The logarithmic slopes of the piecewise power-law fits, $\Gamma$, and the mass range applicable to each segment are summarized in Table \ref{Table1}. This table also contains (normalized to one solar mass) the number of ionizing photons and the total stellar luminosity at age 1~Myr. In the final column of the Table, we give the mass fraction in the stars with masses $> 10M_{\odot}$.

The division between the high-mass and the low-mass stars in these IMFs is of fundamental importance in the determination of star formation rates using H$\alpha$ flux (or any other recombination line). As a consequence of the similar slopes in the upper IMF,  the number of ionizing photons, and hence the H$\alpha$ flux per solar mass of \emph{massive stars} is very similar. For a stellar population with age 1~Myr and having solar abundance, each solar mass of massive stars will produce an H$\alpha$ flux of $\log F_{H\alpha} = 35.53, 35.55$ and 35.55 for the Salpeter, Miller-Scalo and Kroupa IMF, respectively. However, the H$\alpha$ flux per solar mass of \emph{all stars} is quite different; the corresponding figures are now $\log F_{H\alpha} = 34.63, 35.27$ and 34.84 for the Salpeter, Miller-Scalo and Kroupa IMF, respectively. This is entirely a consequence of the vastly different mass fractions of the low mass stars, which produce very little luminosity, but which account for the majority of the mass. In this sense, the low mass stars act as the low stellar mass tail wagging the luminous, high stellar mass dog. 

As if this is not bad enough, star formation rates inferred from H$\alpha$ are also fairly strongly affected by the metallicity in the star and its surrounding ISM. First, higher metallicity reduces the number of ionizing photons which escape to interstellar space. For our $10^4 M_{\odot}$ cluster at an age of 1.5 Myr, the ratio of the photons which escape the atmosphere relative to the solar case are 1.26, 1.18, 1.07,  and 0.71 for metallicities of $Z=0.05Z_{\odot}, 0.2Z_{\odot}, 0.4Z_{\odot}$, and $2Z_{\odot}$, respectively. The main reason for the reduced photon output at higher metallicity is the shift of the stellar zero-age-main-sequence to lower temperature with increasing Z. For instance, in the stellar evolution models used in \emph{Starburst 99}, a 1.5 Myr old $60 M_{\odot}$ star with $Z=0.05Z_{\odot}$ is $\sim17$  \% hotter  relative to the solar metallicity case. This effect is somewhat (but not fully) compensated by the lower luminosity at lower Z: the same star is 6 \% less luminous at lower Z. In contrast, metallicity has almost no influence on the output of hydrogen ionizing photons by an atmosphere of given effective temperature. Higher metallicity decreases the emergent flux at those wavelengths where the opacity is high, such at the location of spectral lines. However, there is the counteracting back-warming effect of the line-blanketing on the adjacent continuum, which raises the flux emitted in the continuum windows with reduced line opacity. \citet{Kudritzki02} demonstrated that the two effects of blocking and blanketing almost entirely cancel each other, a result also found by \citet{Smith02}.

A second metallicity-dependent factor which influences star formation rate estimates is the competition by the dust in the \HII\ region for the ionizing photons. This was quantified by \citet{Dopita03} who found that the ratio of the recombination line flux with and without dust is simply $f=F_{{\rm H}\alpha}/F_{{\rm H}\alpha}(0)=x_d/x_o$ with
\begin{equation}
{f={\frac{F_{{\rm H}\beta}}{F_{{\rm
H}\alpha}(0)}=y^{-1}\ln \left[\frac{1}{1+y}\right]}},  \label{Abs}
\end{equation}
where $y=(c/{\alpha}(T_e){\cal U}\kappa$, $\alpha (T_e)$ being the effective recombination coefficient at the mean electron temperature of the nebula, and $\kappa$ being the dust opacity. Because the dust opacity scales as the metallicity  it follows that the fraction of EUV photons absorbed by the dust is a function only of the product of the metallicity, $Z$, or (equivalently) (O/H), and the initial ionization parameter, ${\cal U}$. Since this factor depends upon the (flux weighted) distribution of the ionization parameters in the ensemble of \HII\ regions, its importance in star formation rate determinations is more complex to evaluate, but can easily be $\sim 30$\%.

\clearpage
\begin{deluxetable}{lcccccc} 
\tablecaption{Initial Mass Function Data \label{Table1}}
\tablehead{
\colhead{Form of IMF:} & \colhead{$\Gamma$}& \colhead{$M_{\rm low}$} & \colhead{$M_{\rm up}$} & \colhead{$\log Q_{\rm H}$} & \colhead{$\log(L$/erg~s$^{-1}$)} & \colhead{$f(M>{10}M_{\odot})$}\\
}
\startdata
Salpeter: & ~-2.35 & ~0.1 & ~120 & 46.49 & 36.42 & 0.127\\
Miller-Scalo: & -0.4 & ~0.1 & 1.0 & 47.13 & 37.05 & 0.527\\ 
 & -1.5 & ~1.0 & 10.0 \\ 
& -2.3 & ~10.0 & 120 \\ 
Kroupa: &  -1.3 & 0.1 & 0.5 & 46.70 & 36.62 & 0.196\\ 
& -2.3 & 0.5 & 120 \\ 
\enddata
\end{deluxetable}
\clearpage

We conclude that metallicity of the exciting stars, or the effects of dust absorption in the ionized plasma  can each affect estimates of star formation rates based upon the H$\alpha$ luminosity by $\sim 30$\%. However,  much larger errors, of order a factor of two, can be made if we do not possess a good model of the low mass end of the IMF. 

Table 1 also reveals that the total luminosity per unit star formation is different between different models of the IMF, and this will directly affect estimates of total star formation rates based upon the 8-1000~$\mu $m infrared luminosity, $L_{{\rm IR}}$. Generally, this estimate is made by assuming that the dusty shell is entirely wrapped around the star formation region so that the IR emission acts like a stellar bolometer. With this assumption, for a stellar population with age 1~Myr and having solar abundance, each solar mass of \emph{massive stars} will produce an IR flux of $\log \left [L_{\rm IR}/{\rm erg s^{-1}} \right ] = 47.39, 47.41$ and 47.41 for the Salpeter, Miller-Scalo and Kroupa IMF, respectively. However, the IR flux per solar mass of \emph{all stars} is again quite different; the corresponding figures are now  $\log \left [L_{\rm IR}/{\rm erg s^{-1}} \right ] = 46.49, 47.13$ and 46.70 for the Salpeter, Miller-Scalo and Kroupa IMF, respectively. Again, the assumptions about the form of the IMF \emph{alone} introduce potential uncertainty in inferred star formation rates of more than a factor of two! These factors are of great concern in attempts to infer the star formation history of the Universe, and deserve to be studied in greater depth.

\section{The Molecular Gas Dissipation Timescale}
In Paper I, we demonstrated that not only is the pressure of the ISM important in determining the form of the spectral energy distribution (SED) from \HII\ regions, but also the amount and the distribution of the molecular gas around the ionized region. This determines both blocking factors at optical-UV wavelengths, and dust temperature distributions which influence the far-IR spectrum. We introduced the concept of a molecular cloud dissipation timescale to quantify the importance of both the obscuration by dark molecular clouds, and the importance of dust re-emission in the molecular gas in determining the overall form of the SED. In our model, the solid angular covering factor of molecular gas about the \HII\ region was assumed to decline exponentially with a time-constant which we called the molecular gas dissipation timescale, primarily determined by the shredding of the placental molecular cloud by the expansion of the \HII\ region.

If we are interested in using H$\alpha$ emission in order to determine star formation rates, or else to understand the nature of the exciting clusters of \HII\ regions, we need to understand what fraction of the emitted H$\alpha$ flux is being blocked by molecular clouds along the line of sight, and therefore to have an estimate of the molecular gas dissipation timescale. Without recombination line observations in the near-IR, or without radio continuum observations, the correction factor due to obscuration by molecular clouds is much more uncertain than the attenuation of the H$\alpha$ emission by a diffuse foreground screen, which can be readily corrected for, once the Balmer Decrement is known. 

Amongst the young, bright and partially dust enshrouded \HII\ regions in our galaxy which are susceptible to detailed analysis, the bright galactic HII region NGC6618 (M17) is a particularly valuable example. This \HII\ region is very young, and has high surface brightness. It is somewhat typical of the bright HII regions we observe in nearby galaxies. Furthermore, it is excited by a cluster which is massive (over a thousand solar masses in total), and its \HII\ region is currently breaking free of its parent molecular cloud. Therefore, an estimate of its age would help constrain a sensible choice of the molecular gas dissipation timescale to be used in subsequent theoretical modeling.

The cluster age can be estimated from the H-R Diagram of the 10 O3V to O9V stars which it contains.  Because the cluster is heavily obscured by overlying molecular gas having between 3 and 15 magnitude of visual extinction, it is best seen in the infra-red, as in the recent Subaru Telescope IR imaging observations \citep{Jiang2002}.  

We have combined multi-band photometry from both optical and mid--infrared observational data as given by \citet{Hanson1997, Ogura1976, Chini1980} and \citet{Chini1998}. The stellar evolutionary tracks are taken from \citet{Schaller1992}. In a similar manner to \citet{Hanson1997}, the stellar class -  $T_{\rm eff}$ calibrations were used to derive a common solution to both the extinction and distance modulus. Using the 11  stars in Table \ref{Table2}, an average spectroscopic / extinction distance modulus, $DM = 10.70$, $d = 1.9$~kpc was obtained. In this, we have made allowance for the star B189 which is most likely a multiple star - possibly a binary O5 or O4 star.  The distance estimate  derived here lies between the  $DM = 10.5$ ( $d = 1.6$~kpc) found by \citet{Hanson1997} and $d = 2.2$~kpc ($DM = 10.85$) given by \citet{Chini1998}. This is simply a result of our averaging of the combined photometric data sets.  This compared favourably with the recent \emph{Hipparcos} distance determination of 1.814~kpc \citep{Kharchenko2005}.

\begin{table} 
\begin{center} {\small \caption{M17 O-Star HR-Diagram data: DM = 10.70 (1.9kpc)}}
\begin{tabular} {l l r r r}
\hline
\multicolumn{1}{c}{HHC97 } &
\multicolumn{1}{c}{Spectral}&
\multicolumn{1}{c}{$\log[T_{\rm eff}]$}&
\multicolumn{1}{c}{$\log[L_{\rm bol}]$}&
\multicolumn{1}{c}{$\log[L_{\rm bol}]$}\\
\multicolumn{1}{c}{No.} 
&\multicolumn{1}{c}{Type$^a$}
&\multicolumn{1}{c}{(K)}
&\multicolumn{1}{c}{$L_{\odot}$}
&\multicolumn{1}{c}{ (ZAMS)$^b$}\\
\hline
\hline
B98 	&09V	&4.56	&4.99	&4.73\\
B111	&05V	&4.65	&5.40	&5.47\\
B137	&O4V	&4.67	&5.69	&5.65\\
B164	&O7.5V	&4.59	&5.05	&5.02\\
B174	&O3V	&4.69	&5.90	&5.82\\
B181	&O9.5V	&4.54	&4.55	&4.64\\
B189$^c$	&O4/5	&4.65	&5.92	&5.47\\
B260	&O7.5V	&4.59	&5.21	&5.02\\
B289	&O9.5V	&4.54	&4.66	&4.64\\
B311	&O9.5	&4.54	&4.59	&4.64\\
OI345$^d$	&O6	&4.63	&5.31	&5.29\\
\hline
\multicolumn{5}{l}{$^a$ HHC97 classification.}\\
\multicolumn{5}{l}{$^b$ From interpolated tracks.}\\
\multicolumn{5}{l}{$^c$ Very likely a binary star}\\
\multicolumn{5}{l}{$^d$ Based on OI76 UBV data.}\\
\end{tabular}
\end{center}
\end{table}

The resulting H-R diagram is shown in Figure \ref{fig1} for the adopted distance modulus, $DM=10.7$. An cluster age of $0-2$~Myr is deduced from this plot.  If we choose to minimize the number of stars which fall below the Zero-age Main Sequence on this plot, we would then require a larger distance modulus;  $DM = 10.75$ which implies a distance somewhat greater than 2.0~kpc. However, in this case the inferred cluster age spread remains the same, around $2$~Myr.  This larger distance would be somewhat incompatible with the \emph{Hipparcos} estimate. In a completely different approach using the incidence of disks around the central stars, \citet{Jiang2002} infer a maximum cluster age of 3~Myr, but they suggest that it may be much younger because disks around massive stars are known to dissipate more quickly. 

We conclude that the probable age of the cluster is $1.0^{+1.5}_{-1.0}$~Myr, and therefore that the use of a molecular cloud dissipation timescale of 1~Myr is appropriate for \HII\ regions. In the adoption of this number, we must recognize the danger of relying upon one object to provide a calibration, but we are hindered by the scarcity of nearby \HII\ regions of appropriate luminosity and evolutionary status. 

Various other ways of estimating this quantity can be developed. The first to to compare the positions of bright extragalactic \HII\ regions with theoretical isochrones on line ratio diagnostic diagrams. This will be done in the next paper of this series. A second and very direct technique would be to compare the emission line flux coming from heavily dust embedded \HII\ regions in the IR with that seen at optical wavelengths. As can be seen in figure \ref{fig2}, below, the fraction of the total number of ionizing photons emitted is a very sensitive function of age. Therefore, averaged across a galaxy, the measured optical to IR line flux ratio is also very sensitive to the molecular gas dissipation timescale. A particular example of this method is provided by the ratio of the 3868\AA\ [\ion{Ne}{3}] line and the 15.55\mum\ [\ion{Ne}{3}] line in the IR. Assuming that the 3868\AA line can be corrected for general interstellar reddening and that the dust absorption in the 15.55\mum\ {\ion{Ne}{3}] line is negligible, for solar abundance the intrinsic time averaged 3868\AA\ [\ion{Ne}{3}] /15.55\mum\ {\ion{Ne}{3}] line ratio is 0.217, 0.139, 0.095 and 0.072 for a molecular cloud dissipation timescale of 0, 1, 2 and 3 Myr, respectively. Projects such as SINGS Survey \citep{Smith04,Regan04} will provide very useful data in helping to observationally constrain the molecular gas dissipation timescale through this method.

\clearpage
 \begin{figure} 
 \begin{center} 
 \includegraphics[]{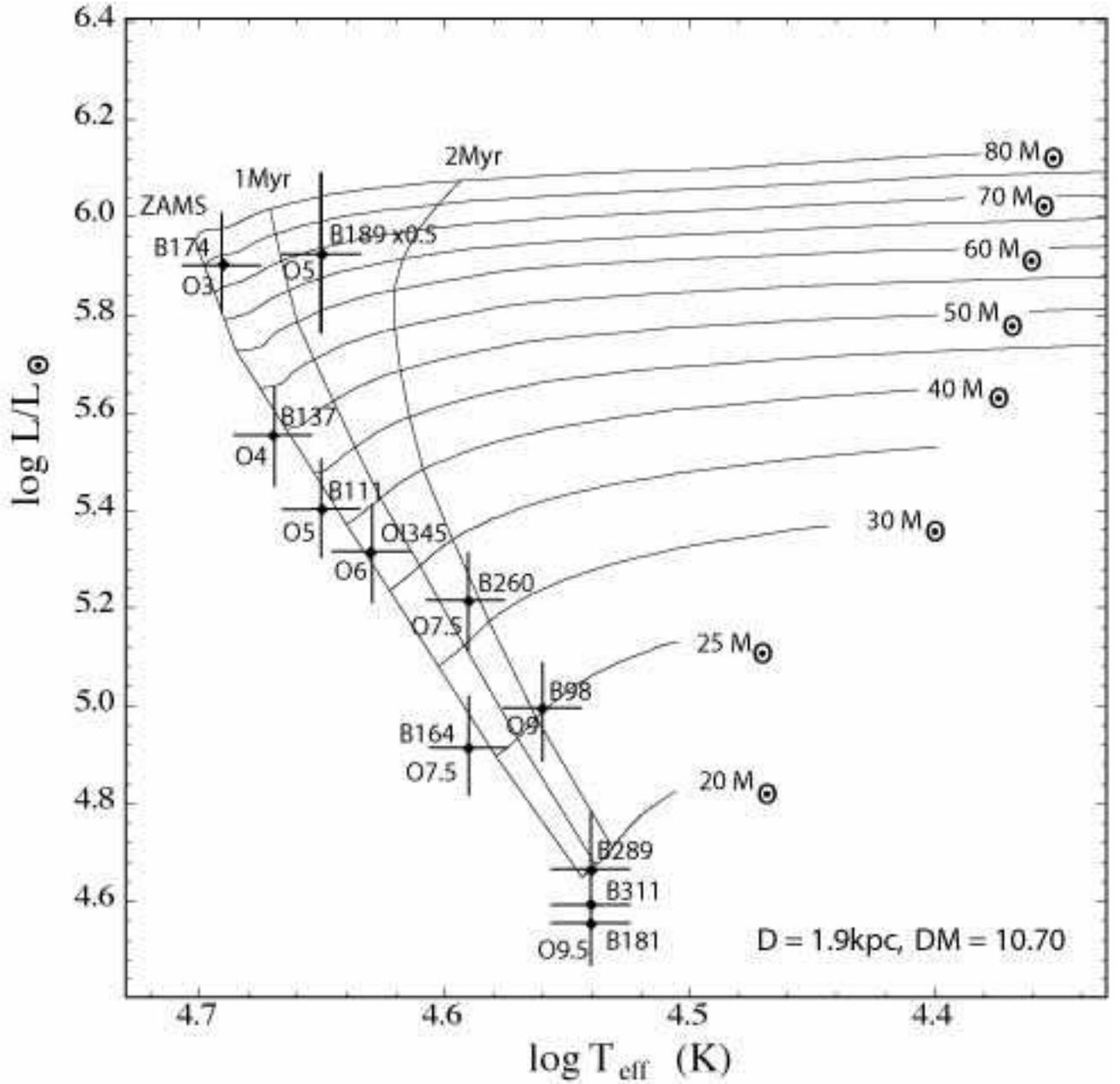}
  \caption{\label{fig1} A $L-T_{\rm eff}$ Hertzsprung-Russell Diagram for NGC6618 (M17).  The indicated error bars simply represent the photometric uncertainties. The age of the cluster implied by this diagram is $1\pm1$~Myr . \label{Table2}}
  \end{center} 
  \end{figure}
\clearpage

\section{The Cluster Mass Function}

The distribution of of stars (by number) in young clusters has been established by observations of objects in our Galaxy, in the Magellanic Clouds  and in other nearby galaxies to follow a power law form \citep{Hunter03, Zhang99, Meurer95};
\begin{equation}
N(N_*)dN_* \propto N_*^{-2}dN_*. \label{Number}
\end{equation}

Most recently, the comprehensive study of the SMC cluster population by \citet{Oey04} has shown that this relationship appears to continue all the way down to $N_* \sim 1$.This result is consistent with the idea that there exists both a universal IMF and a universal $N_*^{-2}$ clustering law.

Within our Galaxy, \citet{Lada03} has reviewed the evidence relating to the mass function of embedded clusters in active star formation regions. For masses in the range 100-1000~M$_{\odot}$ the cluster mass function can be represented by:
\begin{equation}
d(N_*)/dM_* \propto M_*^{-2}.  \label{Number}
\end{equation}
Below $\sim100$~(M$_{\odot}$ the cluster mass function falls steeply, with a shape consistent with a lognormal distribution.  \citet{Lada03} were unable to establish the shape of the IMF for cluster masses in excess of 1000~M$_{\odot}$, because of the scarcity of massive Galactic clusters.

The form of the high mass end of the cluster IMF is better established by more indirect means such as measurement of the H$\alpha$ luminosity function of \HII\ regions, especially of Giant \HII\  regions in external galaxies. The quantity being measured by this means is not simply the number of exciting stars per cluster, but is more closely related to the actual mass distribution of the stars in the cluster through the stellar luminosity:mass relation.

Like the cluster star number distribution, the \HII\ luminosity distribution has usually given as a power-law fit \citep{Kennicutt89,Wyder97,Youngblood99,Oey03}, at least as far as its upper section is concerned. However, many observations show a curvature of the slope of the luminosity function to lower values (a flattening) at lower luminosities, as first pointed out by \citet{Kennicutt89}. This flattening does not arise solely from incompleteness. Indeed, in the case of M33 \citep{Wyder97} where observations extend to very low \HII\ region luminosities, and where corrections sample completeness have been carefully made, the number counts turn over at an H$\alpha$ luminosity of $10^{36.4}$ erg~s$^{-1}$. A similar turnover was found for irregular galaxies by \citet{Youngblood99}. This turnover has been ascribed by Oey and her co-workers \citep{Oey98,Oey03} as due to stochastic variation in the number of ionizing stars in low-mass clusters. For these ``sparse" clusters, the ionizing flux can vary widely, leading to a much greater scatter in the H$\alpha$ flux of their \HII\ regions.

While the stochastic variations in ``sparse" clusters are certainly important, another, possibly more important effect remains so far neglected; the decrease in the ionizing flux of the cluster with the age of the central cluster. We will now consider this effect, and its dependence upon metallicity.

Because the lifetime of the most massive stars which produce the majority of ionizing photons is so limited, the ionizing photons are emitted within a short period after the birth of the cluster. In figure \ref{fig2} we show the fraction of ionizing photons as a function of time and metallicity. In order to correct for the compact \HII\ region phase, when the central star cluster is still dust-embedded, we have used a molecular cloud dissipation timescale $\tau$, introduced in Paper 1 and discussed in the previous section, of 1~Myr. The fraction of the observable H$\alpha$ in the surrounding \HII\ region is then $(1 - exp[-t/{\rm Myr}])$. With this form, we compute that  $\sim 50$\% of the integral H$\alpha$ luminosity is produced by between 1.5 and 2.5 Myr, although the exact percentage depends somewhat upon the metallicity of the cluster.

As long as the most massive stars of the cluster remain on or near the Zero-Age Main Sequence (ZAMS), the production rate of ionizing photons remains approximately constant, but then rapidly declines as the most massive stars expire in supernova explosions. Figure \ref{fig3} shows the probability distribution of the H$\alpha$ luminosity of a $10^4M_{\odot}$ cluster of solar metallicity integrated though all cluster ages for which there is an appreciable production of ionizing photons. The distribution shows a well-defined peak produced by clusters between 0.5 to 2.5 Myr, and an extended tail towards lower luminosities due to photons emitted at later times.

In terms of the probability of encountering a cluster at a given luminosity, the distribution is much flatter, since the older clusters take longer to fade though 0.1dex of luminosity. We therefore expect to find a large population of faint, large and evolved \HII\ regions. However, since these will have both low H$\alpha$ flux and low emission measure as a consequence of their large sizes and low pressures, they may be difficult to observe independently, although they may contribute as much as 30\% of the total recombination line luminosity of a galaxy. We discuss these objects in more detail in the following section.

The observed luminosity function of the \HII\ regions in a galaxy is the convolution of the metallicity and time dependent luminosity of the \HII\ region surrounding a given cluster and the mass function of the clusters. As long as this mass function does not persist as a power-law to low masses, a break in the slope of the luminosity function wil be produced by the combination of the shape of the lower mass cut in the cluster mass function, the fading of the clusters at late time, and any stochastic stellar mass variation.

\clearpage

\begin{figure}
\includegraphics[]{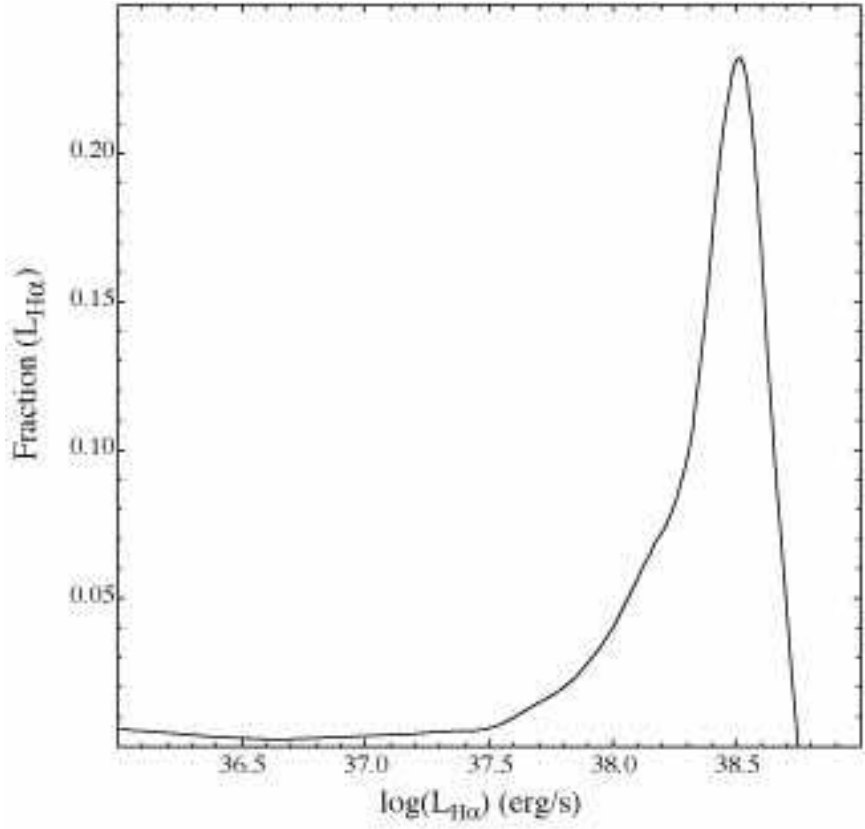}
  \caption{\label{fig2}
  The fraction of ionizing photons emitted as a function of time and stellar metallicity. Note that virtually all the ionizing photons have been emitted by 10 Myr.}
\end{figure}
\clearpage

\clearpage
\begin{figure}
\includegraphics[]{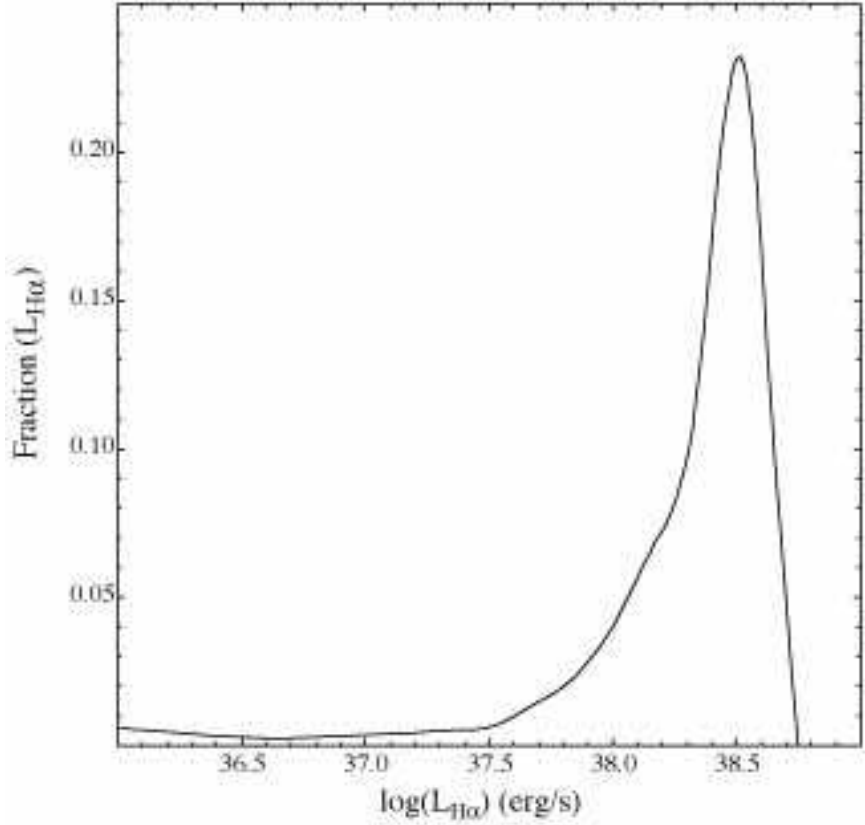}
  \caption{\label{fig3}
  The fraction of the total luminosity in H$\alpha$ of a $10^4M_{\odot}$ cluster emitted per 0.1dex of H$\alpha$ luminosity as a function of the H$\alpha$ luminosity of the \HII\ region. This shows that while most of the luminosity is emitted when the cluster is young and at or near the peak luminosity, there is a very extended tail in the H$\alpha$ emission from faint extended  \HII\ regions at late times}
\end{figure}
\clearpage

Although astronomers are fond of using power-law distributions, any multiplicative random process (which may well describe the birth of stellar clusters) will produce a log-normal distribution rather than a power-law. Such a distribution is better-behaved mathematically than the pure power-law form. It may well have an added power-law at its high end. This case is known from economics as a Pareto distribution, where it has been used to characterize such diverse applications as the distribution of incomes in a population, or the movement of stocks or the size of cities.

In the ISM, it is known that isothermal turbulence leads naturally to a log-normal probability gas density function \citep{Padoan97,Scalo98,Ostriker01} which \citet{Padoan02} has shown to be consistent with a power-form of the cloud core mass function given that the power spectrum is a power law, and assuming that the cores of the cloud producing clusters have a size which is comparable to the thickness of the post-shock layers. 

Theoretical models for the stellar mass function also share features of a Pareto distribution. For example \citet{Basu04} and \citet{Bate05} both propose that the opacity limit for fragmentation sets the characteristic low-mass peak in the IMF, which can be described by a log-normal function, but that processes of protostellar ejection and accretion on massive cores can drive shape of the upper end of the IMF into a pure power law.

Here, we assume that all cluster mass functions can be fit by a log-normal distribution in mass $m$ of the form:
\begin{equation}
P(m)= {\frac{{1}}{{\sqrt{2\pi}\sigma m}}} exp\left[ -{\frac{(\ln m - \ln m_0)^2}{2 \sigma^2}}\right]  \label{lognorm}
\end{equation}
where $\ln m_0$ is the mean, and $\sigma$ is the variance of $\ln m$. The initial luminosity of the cluster is assumed to be proportional to the cluster mass, and the \HII\ luminosity function is constructed by convolving the cluster mass function with the fading function of the clusters. This provides an extended flat region below a luminosity corresponding to the peak of the cluster mass function; $m_0$.

We have fitted to a sub-sample of observed luminosity functions which cover the broadest range, and are reasonably complete down to the lower luminosities ($logL\sim 37$) needed to adequately fit for $m_0$. This sample is M33  \citep{Wyder97}; the \emph{Hubble Space Telescope} data for M51 \citep{Scoville01}, the ground-based data for M100 \citep{Oey03} and the sum of those Im and SM galaxies which have distances of $<3.5$Mpc from  \citet{Kennicutt89}. This defines a composite luminosity function for  the LMC, SMC, NGC2366 and NGC4449, necessary in order to ensure adequate counting statistics. Only the M33 data has been corrected for incompleteness at low luminosity. For the remainder, the data are terminated at a luminosity below which incompleteness effects become clearly evident.

\citet{Scoville01} have emphasized the importance of adequate spatial resolution in determining the luminosity function, especially when this is a steeply falling function of luminosity. The small \HII\ regions become lost in vicinity of the more luminous objects, and the resulting composite is counted as one more luminous \HII\ region. This causes the observed luminosity function to flatten over a broader range of luminosities. This effect has the potential to radically alter the fitting parameters. To avoid this, we have only selected galaxies which have observations of adequate spatial resolution, with the possible exception of M100.

Another factor which can alter the fitting parameters is the foreground extinction at H$\alpha$, which \citet{Scoville01} have demonstrated to be quite important in the case of M51. We have no way of correcting for this with the current data sets.

The log-normal fits are shown in figure \ref{fig4}. The error bars shown on the data are purely the $\sqrt N$ counting statistics. The oscillations which appear in the theoretical luminosity function at low H$\alpha$ luminosities are a numerical artefact. They result from of our binning of the probability function of the cluster ionizing luminosity into 0.1~dex bins, and from the 0.1~Myr steplength used in the 
Starburst 99 models. At certain times the cluster fades rapidly, and therefore the function which convolves the mass function to generate the H$\alpha$ luminosity function is not entirely smooth. Overall, a very good fit is achieved for all four luminosity functions. Table \ref{Table3} gives the fitting parameters of the log-normal mass function. 

Amongst these galaxies, the variance in the cluster mass function shows quite a large spread from one galaxy to another, which makes a large difference at the upper mass end of the mass function. However, for theoretical work, it is probably adequate to take the mean of the fitting parameters for the cluster mass function; $m_0 = 100 M_{\odot}$ and $\sigma = 1.7$. These values will be adopted in the remainder of this paper.

\clearpage

\begin{figure*}
\includegraphics[]{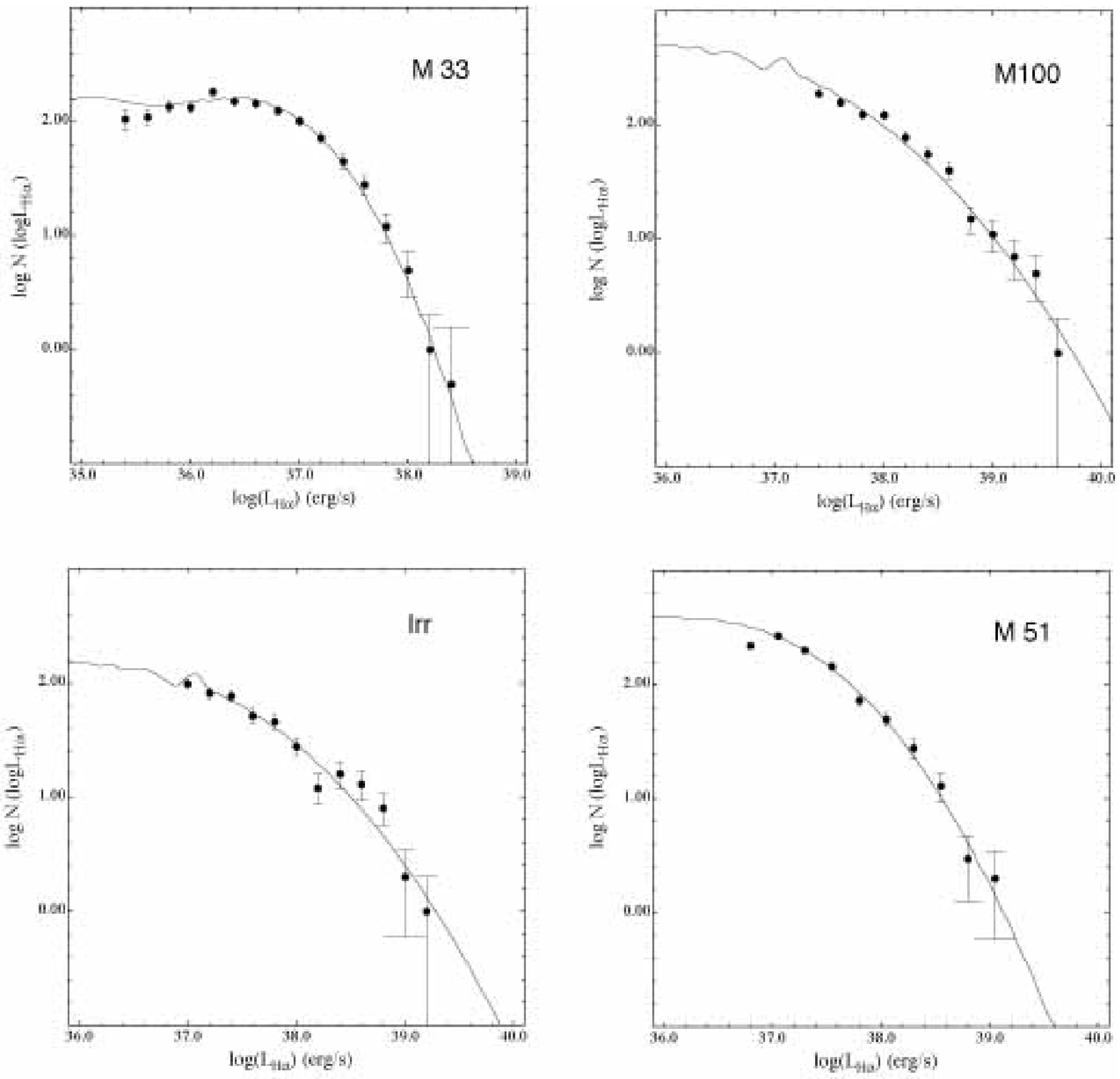}
  \caption{\label{fig4}
  Fits showing the variety of observed differential \HII\ region H$\alpha$ luminosity functions. The observations are drawn from  \citet{Kennicutt89,Scoville01} and \citet{Wyder97}. The error bars are the $\sqrt N$ statistical errors. Only the M33 data has been corrected for incompleteness. For the others, only those portions of the luminosity function in which the data are reasonably complete are shown. The fitted luminosity functions are shown as lines. All of these are consistent with a log normal form for the cluster mass function. In all cases, the fit is tighter than would have been obtained using a truncated power-law form.}
\end{figure*}

\clearpage
\begin{deluxetable}{ccc} 
\tablecaption{Fitting parameters for the cluster mass function \label{Table3}}
\tablehead{
\colhead{Galaxy:} & \colhead{$m_0/M_{\odot}$}& {$\sigma$}  \\
}
\startdata
M33 & 95 & 1.14\\
M51 & 95 & 1.63\\ 
M100 & 95 & 2.12\\ 
Irr & 120 & 1.95 \\ 
\enddata
\end{deluxetable}
\clearpage

The effect of limited resolution is graphically estimated from comparing the fit obtained for M51 using the \citet{Scoville01} space-based data with the ground-based data from \citet{Oey03}. Although both give a similarly good fit to a log-normal function and give the same value of $\sigma$, 1.63,  a value of $m_0 \sim 1400 M_{\odot}$ would be estimated using only the ground-based data. This is the result of the ``flattening" of the luminosity function referred to above.

Stochastic variations in the cluster IMF at low cluster mass \citep{Oey97,Oey98} are certainly important. The effect of these is to provide a further apparent broadening of the cluster H$\alpha$ luminosity function at low mass. However, as far as the H$\alpha$ luminosity function is concerned, this effect will be largely masked masked by the fading of the more massive clusters with time, since these dominate over the lower mass clusters at the lowest H$\alpha$ luminosities. The `tail' of fading massive clusters is the cause of the flat luminosity function fits at low luminosities seen in figure \ref{fig4}. For example, in the M51 fit, at $log(L_{H\alpha})=36$ erg~s$^{-1}$, 48\% of the clusters will have masses greater than $10^3 M_{\odot}$ and 65\% have masses greater than $400 M_{\odot}$. Nonetheless, both fading and stochastic variation need to be taken into account when fitting the lower end of the luminosity function. Qualitatively, the effect of stochastic variation will be to cause us to overestimate $m_o$ somewhat.

It is noteworthy that all galaxies have $m_0 \sim 100 M_{\odot}$. Such a low value of  $m_0$ is at first sight a little surprising, since the most probable cluster mass is similar to the upper cutoff in the stellar IMF. However, it is consistent with the observation by \citet{Oey04} that the cluster OB star number relationship for clusters appears to continue all the way down to the field stars (\emph{i.e.} ``clusters" composed of single stars).  It is also consistent with the shape of the embedded cluster mass function at the low mass end as given by \cite{Lada03} for those molecular clouds for which observations are reasonably complete to low mass (L1630, L1641, Perseus and Mon R2).

In effect, this low value of $m_0$ ensures that, in the mass range where the stellar IMF is falling, the cluster mass function is rising. The cluster mass function reaches its maximum at about the mass where the stellar mass function effectively terminates. We could therefore regard the cluster mass function as the continuation of the stellar mass function ``by other means". Presumably, collapsing clouds cease to form single stars once the mass or the angular momentum of the parent molecular cloud becomes too large. This idea would be consistent with what we think we understand about gravitational instability in a fractal, turbulent medium \citep{Elmegreen04}.

\section{Control of the Ionization Parameter}
We used the stellar wind and supernova power output for clusters having a Miller-Scalo initial IMF as predicted by the \emph{Starburst 99} code to solve for the radius of the \HII\ regions as a function of time. For this, we made a  Runge-Kutta integration of the equation of  motion of the \HII\ shell, in the same manner as was done in Paper I. We briefly summarize the method used here.

For a time-dependent wind + supernova mechanical luminosity $L_{\rm mech}(t)$ and an ambient medium density of $\rho_0$, the equations of conservation of mass, momentum and energy for the mass $M$ of the swept-up shell of shocked ISM gas are; 
\begin{eqnarray}
\frac{d}{dt}[ M ] & =  & 4 \pi R^2  \rho_0 \dot{R} , \\
\frac{d}{dt}[ M \dot{R} ] & =  & 4 \pi P R^2 , \\
\frac{d}{dt}[ P R^3] & =  & \frac{L_{\rm mech}(t)}{2 \pi} - 2 P R^2 \dot{R} , 
\end{eqnarray}
where $R$ is the outer shell radius.  Here $P$ is the bubble pressure, and  $P > P_0$, the ambient ISM pressure. Assuming  $\gamma = 5/3$ defines the equation of state of the bubble gas, we can combine these into a single differential equation in terms of the time-dependent input mechanical luminosity and the density in the surrounding ISM, 
\begin{equation}
 \frac{d}{dt}[R \frac{d}{dt}(R^3 \dot{R})] +
\frac{9}{2} R^2 \dot{R}^3 = \frac{3 L_{\rm mech}(t)}{2 \pi \rho_0}. \label{R-t}
\end{equation}  
This equation was then integrated using a standard Runge-Kutta 4th order integration with an adaptive step size, using the input mechanical energy luminosity data as tabulated by  \emph{Starburst 99}. The internal pressure of the bubble at any radius is given by equation  \ref{P1}. Because the bubble has the form of a mass-loss blown cavity, then much of the interior of the bubble is filled either by the free-wind region, or by hot shocked stellar wind material at coronal temperatures. The \HII\ region proper is confined to exist beyond the contact discontinuity between this shocked wind and the shocked ISM. The contact discontinuity is usually located close to the outer shocked shell of swept-up gas, so we can approximate the inner radius of the \HII\ region by the bubble radius given by equation \ref{R-t}.

Two cases are possible. Either the ionized shell is thin in comparison to the radius of the mass-loss bubble, or else it is thicker and extends into the un-shocked ISM beyond the radius of the bubble. Which case applies depends on the ionization parameter $q$ at the contact discontinuity, which is the inner boundary of the \HII\ region. Generally speaking, the \HII\ region will become thick when log$(q)\geq 8.0-8.5$. The exact value at which this occurs will depend upon the details of the geometry for the particular case considered.

Because the \HII\ region is ionized and is always at a temperature $T_e \sim 10^4$K, its particle density $n$ is determined by the pressure $n=P/kT_e$. Let $\delta(t)=n/n_0$ be the instantaneous ratio of the density interior to the bubble to the density in the surrounding ISM. It therefore follows from  equation \ref{P1} that $n \propto L_{\rm mech}\delta^{-3/2}R^{-2}$. However, if the ionizing photon flux produced by the exciting cluster is $S_*(t)$, then the ionizing flux at the inner boundary of the bubble is $S_*(t)/4\pi R^2$. Therefore the ionization parameter at the contact discontinuity in the photo-ionized plasma depends largely on the instantaneous properties of the exciting cluster stars;
\begin{equation}
q(t) \propto \delta(t)^{3/2}S_*(t)/L_{\rm mech}(t). \label{q}
\end{equation}

The presence of the $\delta(t)$ factor in the above equation provides a weak coupling between the ionization parameter and both the pressure in the ISM, $P_0$, and he mass of the central cluster, $M_{cl}$. Together, these determine the strength of the outer shock of the mass-loss bubble and therefore the compression factor through it. The appropriate scaling factors are  $q \propto P_0^{-1/5}$ and  $q \propto M_{cl}^{1/5}$. 

In figure \ref{fig5} we show the run of ionization parameter as a function of time and ISM pressure for a cluster with $M_{cl}=10^3M_{\odot}$ and a Miller-Scalo IMF. This is computed using the \emph{Starburst 99} output, and solving for the time dependent radius and internal pressure of the mass-loss bubble as described above. In the early phases ($t < 2$~Myr) the ratio of ionizing photon flux to mechanical energy flux is almost constant, and the changes in q mostly reflect the size evolution of the mass-loss bubble. After 2~Myr, the ionizing photon flux rapidly decreases as massive stars evolve to red supergiants, and then the mechanical energy flux increases as a result of Wolf-Rayet stellar winds. Both of these lead to a strong decline in $q$. At around 3.5~Myr, supernova explosions add to the internal pressure of the bubble, sharply decreasing the $q$ in the ionized gas and re-accelerating the expansion of the mass-loss bubble. At later times, the progressive death of the high-mass stars leads to further declines in the ionizing photon flux, but the decline in $q$ is slower because the internal pressure of the mass-loss bubble also declines, partially offsetting the decline in ionizing photon flux.

The ionization parameter is quite strongly dependent upon the chemical abundance. It is driven by two factors. First, at higher abundance, the stellar wind has a higher opacity and absorbs a greater fraction of the ionizing photons, reducing the $q$ in the surrounding \HII\ region. Second, the atmosphere scatters the photons emitted from the photosphere more efficiently when the atmospheric abundances are higher, leading to a greater conversion efficiency from luminous energy flux to mechanical energy flux in the stellar wind base region. This also leads to a diminution of  $q$ in the surrounding \HII\ region. These factors acting together provide a sensitivity to metallicity of  $q \propto Z^{-0.8}$, approximately.

\clearpage

\begin{figure}
\includegraphics[]{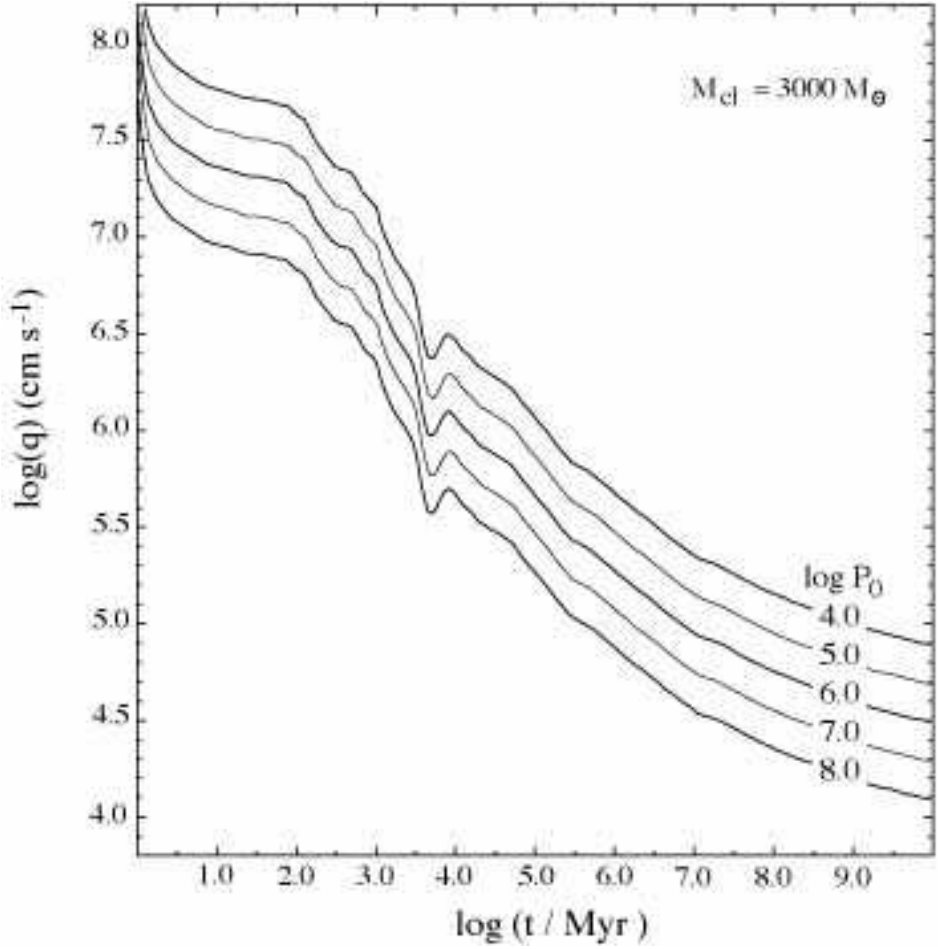}
  \caption{\label{fig5}
The computed run of ionization parameter in the \HII\ region as a function of time and ISM pressure $log P_0/k $ (cm$^{-3}$K) for a cluster having a mass of $10^3M_{\odot}$ and with a Miller-Scalo IMF.}
\end{figure}
\clearpage

\clearpage
\begin{figure*}
\includegraphics[]{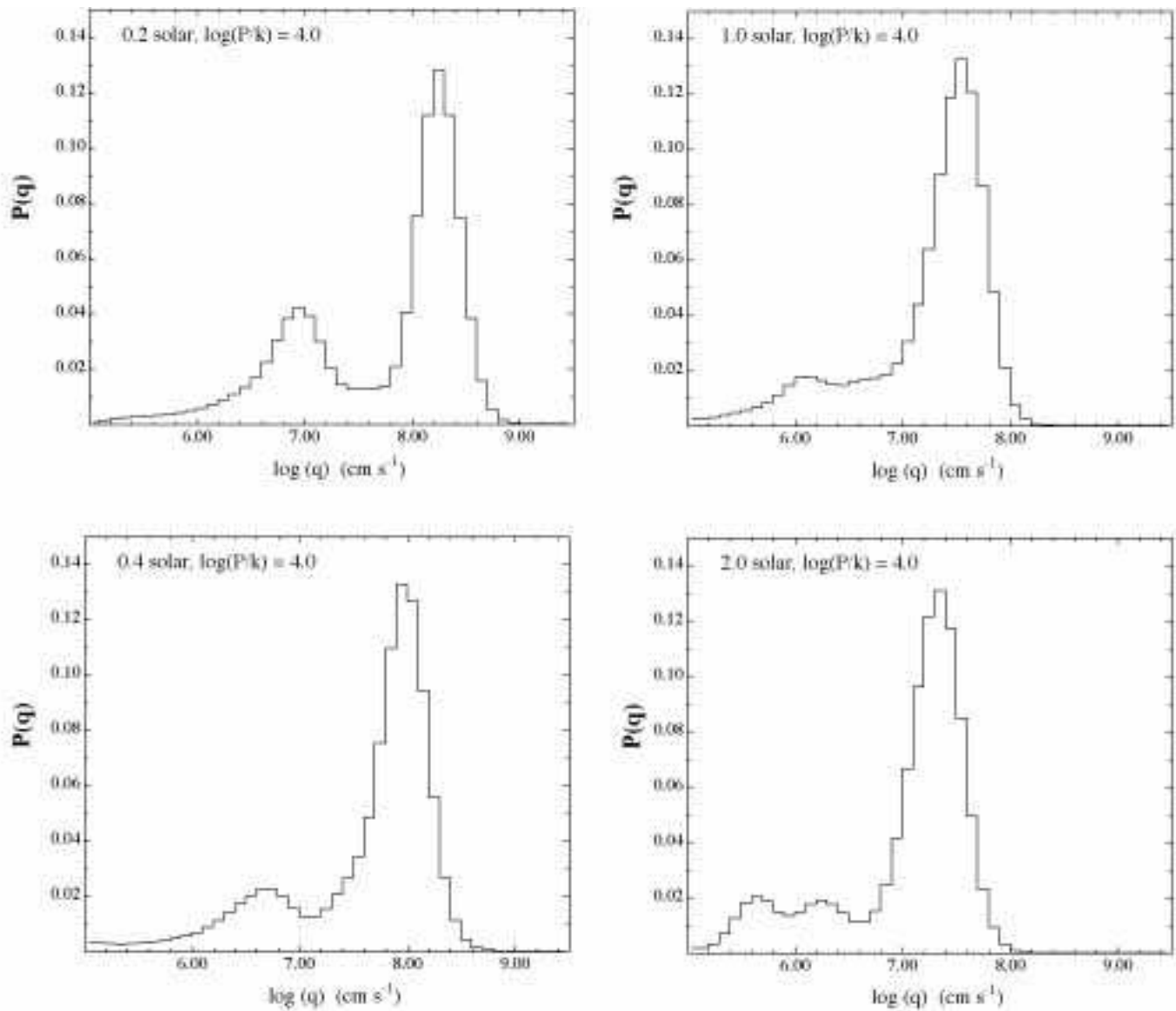}
  \caption{\label{fig6}
The probability distributions of ionization parameter in the \HII\ region as a function of metallicity for a particular ISM pressure $log P_0/k = 4.0$~cm$^{-3}$K. Note that the peak of the distribution is strongly dependent on the metallicity. This also predicts that an important fraction of the \HII\ regions in a galaxy will be evolved large-diameter low emission measure and low ionization parameter objects. These might be identified with the diffuse ionized gas (DIG) component often seen in external galaxies.}
\end{figure*}
\clearpage

A dependence of $q$ on $Z$ had been previously inferred from the observational data, by those attempting to calibrate the abundance scale of extragalactic \HII\ regions using the strong emission line method \citep{Dopita86,Dopita00,Kewley02b}. However, the theoretical reason for such a dependence was not clear, and our work thus provides a solution to this long-standing mystery.

We now have all we need to compute the probability distribution of the ionization parameter at the inner boundary of the \HII\ region as a function of both the chemical abundance and of the ISM pressure in a galaxy. To do this, we assume a Miller-Scalo IMF (although the choice of the exact form of the IMF is rather unimportant in this context), a cluster mass function appropriate to normal galaxies as derived in the previous section; $m_0 = 100M_{\odot}$ and $\sigma = 1.7$, and a molecular cloud dissipation timescale, defined above, of 1.0Myr.

For a given metallicity and pressure $P_0$, this allows us to compute the H$\alpha$ flux-weighted probability distribution of $q$ for an ensemble of \HII\ regions excited by clusters of all ages and masses.
The computed distributions are shown in figure \ref{fig6}. Because the ionization parameter scales precisely as $q \propto P_0^{-1/5}$, we can infer the probability distributions of $q$ at different pressures than $log \left [(P_0/k)/ {\rm cm^{-3}K} \right ] = 4.0$ from figure \ref{fig6}, by applying an appropriate shift in the ordinate.

The major peak in these distributions comes from the young ($0-2.5$Myr) \HII\ regions (\emph{c.f.} fig \ref{fig2}). These \HII\ regions are characterized by high pressures and small sizes, and therefore have high emission measures. These will be the objects preferentially selected for observation, and which dominate the global spectrum of the ensemble of \HII\ regions observed in distant galaxies. In the next paper of this series, we will investigate the emission line abundance and excitation diagnostics of such ensembles of \HII\ regions.

In addition to these bright \HII\ regions, there is a substantial secondary peak and low-$q$ tail of large, evolved objects having low emission measure and low ionization parameter. The secondary peak emerges as a consequence of the onset of the Wolf-Rayet winds and supernova explosions, which cause a high pressure in the ionized gas, and a fairly sudden drop to lower ionization parameters. This behavior is visible in figure \ref{fig5} in the time interval between roughly 3 and 3.5~Myr.

Objects having $\log q < 7.0$ account for between 19\% and 30\% of the total H$\alpha$ emission, depending on the metallicity. Such old \HII\ regions will be characterized by weak [\ion{O}{3}]/H$\beta$ ratios (driven mainly by the lower stellar effective temperatures at late times) and strong [\ion{N}{2}]/H$\alpha$ and [\ion{S}{2}]/H$\alpha$ ratios (driven by the low ionization parameters). These flux fractions and such line ratios precisely characterize those of the diffuse ionized emission (DIM) which is almost ubiquitously seen in disk galaxies \citep{Hoopes96,Martin97,Wang98, Wang99},  and is usually refereed to as the ``Reynolds Layer" in the context of our own Galaxy; see a recent review by \citet{Reynolds04}.  The excitation of the DIM has been usually ascribed to leakage of ionizing photons from young \HII\ regions. However, the emission from large, faint, low pressure and evolved \HII\ regions might provide a better explanation which will be further investigated from the viewpoint of their emission line diagnostics in Paper III of this series.

The mean emission weighted ensemble average ionization parameter is strongly dependent upon metallicity. For $Z/Z_{\odot} = 0.05, 0.2, 0.4, 1.0$ and 2.0, the corresponding $\log \left<q\right> = 8.5, 8.13, 7.90, 7.48$ and 7.28, respectively. All these are given for a pressure of $P_0/k =4.0$~cm$^{-3}$K. As described above, $q$ scales as  $\left< q \right> \propto P_0^{-1/5}$

The above values can be compared with, and used to interpret,  the $\log \left< q \right>$ values inferred from the emission line spectra of galaxies in the \emph{Sloane Digital Sky Survey} (SDSS). The sample of galaxies used here was selected from the full 261054-galaxy SDSS sample \citep{Brinchmann04} according to the following criteria:

\begin{enumerate}

\item A signal-to-noise (S/N) ratio of at least three in the strong emission-lines H$\beta$,  [\ion{O}{3}] $ \lambda 5007$ \AA, H$\alpha$, [\ion{N}{2}]$ \lambda 6584$ \AA, and [\ion{S}{2}]$ \lambda \lambda 6717,31$ \AA.  This S/N  criterion is required in order to allow for accurate classification of the galaxies into either star-formation or AGN-dominated classes \citep{Kewley01a,Veilleux87}.

\item Fiber covering fraction $> 20$\% of the total photometric g'-band light.   \citet{Kewley05} found that a flux covering fraction $>20$\% is required for metallicities to begin to approximate global values.   Lower covering fractions can produce significant discrepancies between fixed-sized aperture
and global metallicity estimates.

\item Stellar mass estimates must be available.  Stellar masses were derived by \citet{Tremonti04} using a combination of $z$-band luminosities and Monte Carlo stellar population synthesis fits to the
D$_{n}(4000)$ spectral index and the stellar Balmer absorption line H$\delta_{A}$
\end{enumerate}

The resulting sample contains 45086 galaxies.  From these, we have removed the  galaxies
containing AGN types using the optical line diagnostics given by \citet{Kewley01} and \citet{Kauffmann03}. For the remainder, the metallicities and the ionbization parameters have been derived  using the methodology described in \citep{Kewley02}. The results  are presented as histograms of $log \left<q \right>$ according to the inferred mass of the parent galaxy in figures \ref{fig7} and \ref{fig8}. 
\clearpage
\begin{figure}
\includegraphics[]{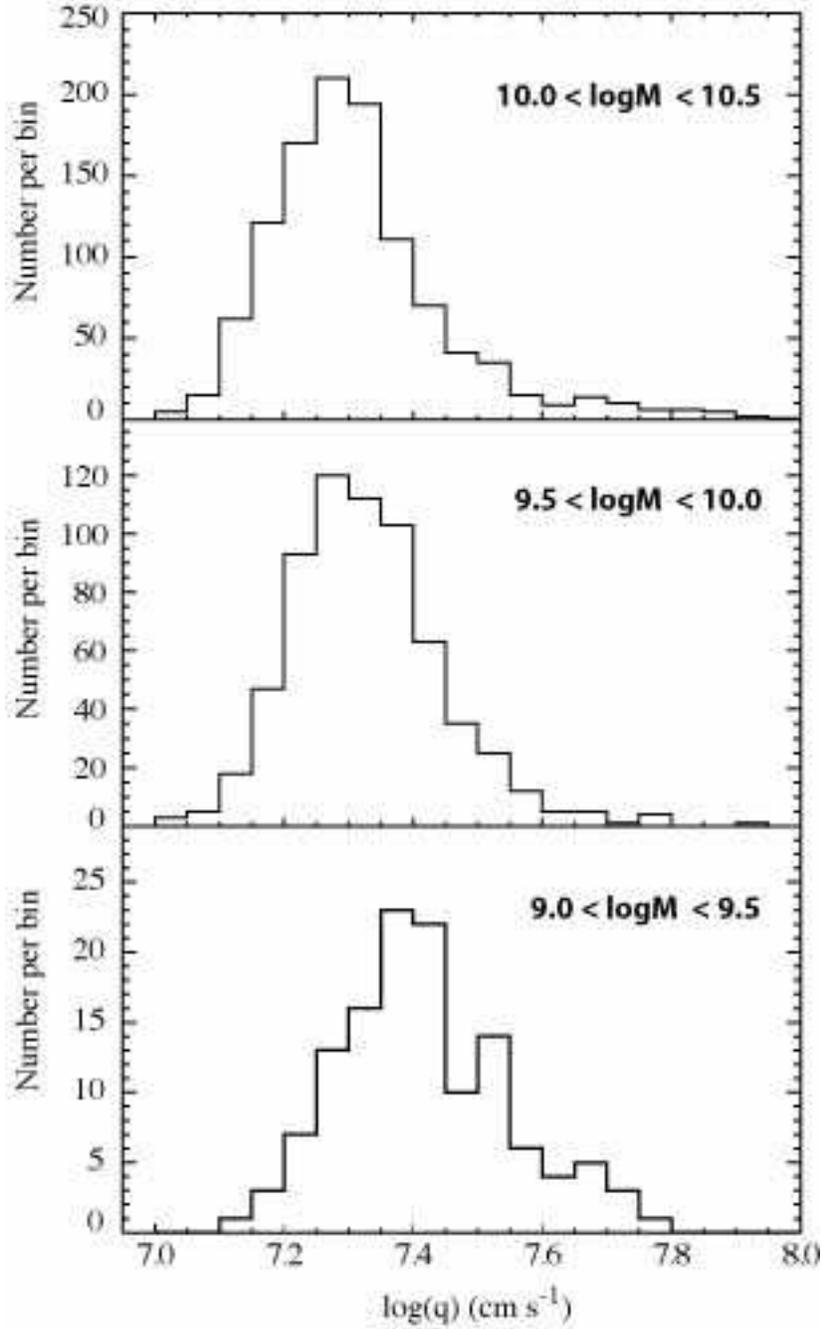}
  \caption{\label{fig7}
The probability distributions of ionization parameter in the SDSS galaxies binned according to the mass of the parent galaxy, as marked.}
\end{figure}

\begin{figure}
\includegraphics[]{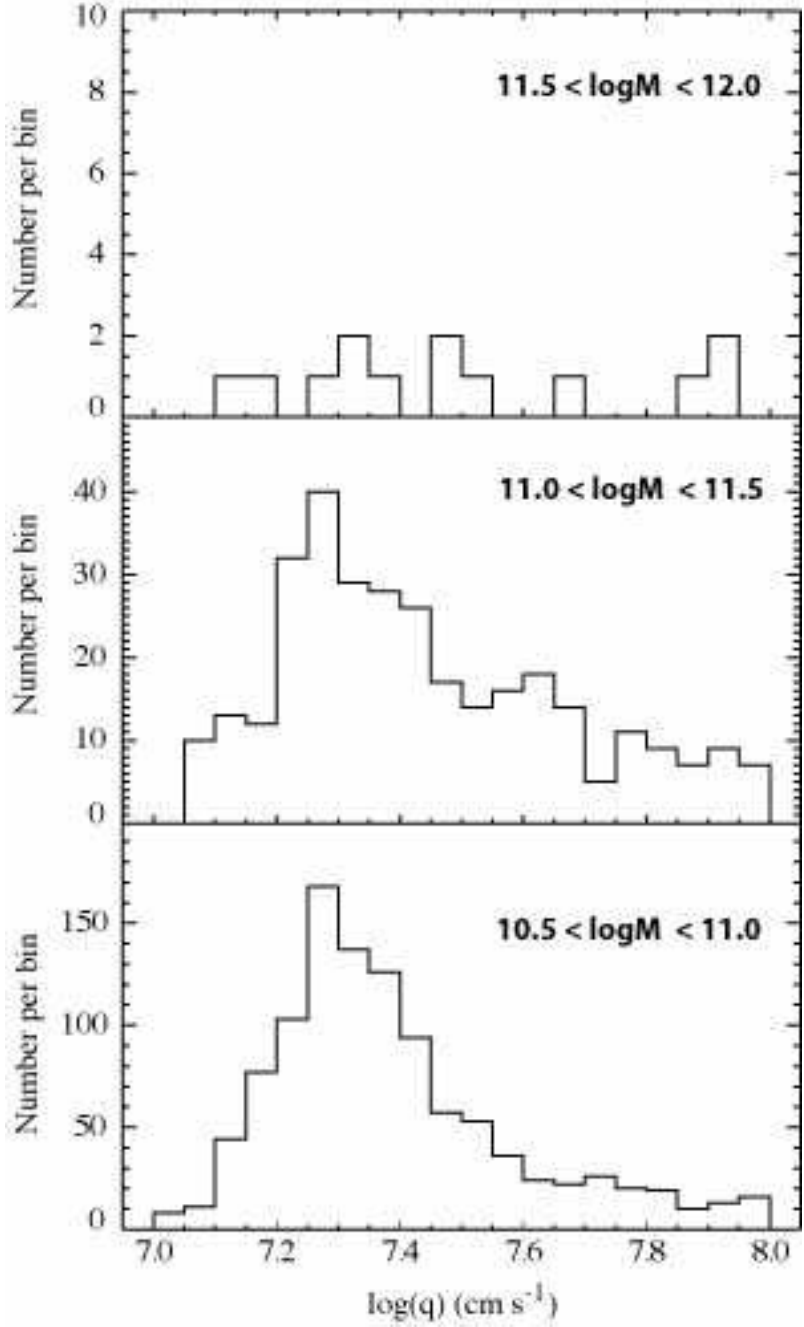}
  \caption{\label{fig8}
As figure \ref{fig7}, but for the most massive SDSS galaxies.}
\end{figure}
\clearpage

Systematic trends in $\log \left< q\right >$ are visible in these figures. As the mass of the parent galaxy increases, the distribution of $\log \left<q \right>$ shifts to lower values. From our theory, this can be ascribed to two effects. First, the lower mass galaxies will have lower metallicities, on average which lead to higher $\log \left<q \right>$. Second, as the mass of the galaxy increases, the depth of the potential becomes greater, allowing higher pressures to be maintained in the ISM of these galaxies. This leads to somewhat lower $\log \left<q \right>$ in the high-mass galaxies. 

In the highest mass bins, there is a resurgence of galaxies having high $\log \left<q \right>$. This is most likely due to the presence of an active nucleus in these objects, which produces a composite \HII\ region plus Narrow Line Region (NLR) spectrum. If present at only a few percent of the flux of the \HII\ region, the AGN contribution to the composite spectrum will simulate an \HII\ region with high $\log \left<q \right>$. This hypothesis needs to be checked by high spatial resolution spectroscopic observations of each of the high $\log \left<q \right>$ galaxies in mass bins above $log M > 10.5 M_{\odot}$.

\section{Discussion \& Conclusions}
\subsection{Sensitivity of Star Formation Rates to the IMF}
Because they are so important in drawing conclusions about the both evolution of galaxies and the global star formation rate throughout cosmic time, we have investigated the sensitivity of commonly-used star formation indicators to assumptions about the IMF. We have found that, even ignoring the possibility that the form of the upper stellar mass part of the IMF may be changed by the local environment (\emph{e.g} truncated or top-heavy IMFs), the uncertainty in star formation rates introduced by assumptions about the form of the lower stellar mass part of the IMF is considerable. For both determinations based upon the H$\alpha$ line, and upon the IR luminosity,  $L_{{\rm IR}}$ errors of order a factor of two arise directly from plausible differences in the assumed form of the lower mass part of the IMF.

\subsection{Cluster Mass Function}
We have demonstrated the the luminosity function of \HII\ regions in galaxies can be fit better by a log-normal cluster mass distribution, rather than a simple power-law, as has been assumed hitherto. The  mean cluster mass is about $m_0 = 100M_{\odot}$ and $\sigma \sim1.7$.  The probability of forming a stellar cluster increases just as the probability of forming a single star decreases, perhaps indicating that cluster formation can be considered as a continuation of the IMF to higher masses. This is probably the result of angular momentum constraints preventing the formation of massive single stars, so that a cluster of stars is formed instead. This conclusion is consistent with those of \citet{Oey04}, based upon number counts of cluster stars in the SMC and the embedded cluster mass function inferred by \citet{Lada03}. This result is also consistent with the idea that it is the density distribution in a turbulent ISM which determines the form of both the stellar and cluster IMF \citep{Elmegreen02,Elmegreen04,Padoan02,Padoan05,Bonnell03}.

\subsection{Ionization Parameter Distribution}
We have demonstrated that the ionization parameter of \HII\ regions, hitherto treated as a free variable, is in fact determined largely by the instantaneous ratio of the ionizing photon flux to the mechanical energy flux of the central stars. Although this conclusion has been obtained using a simple 1-D spherical evolution model for the \HII\ region, it should remain valid in more complex geometries, since wherever un-ionized or molecular inclusions exist, these will be pressure confined by the hot shocked stellar wind gas (modulus the recoil momentum of the gas streaming off the ionization fronts), and this gas pressure is coupled with the overall size of the \HII\ region, which itself determines the dilution of the radiation field of the central stars. Because the ratio of the the ionizing photon flux to the mechanical energy flux of the central stars is a strong function of metallicity according to the stellar atmosphere and stellar wind theory of massive stars, the ionization parameter should also be quite sensitive to metallicity. This appears to be the case for the SDSS galaxies. For individual \HII\ regions there is also some indication of such a systematic variation of ionization parameter with metallicity (\emph{c.f.} figure 6 of \citet{Dopita00}). The ionization parameter is only weakly dependent upon pressure in the interstellar medium and upon cluster mass; $\log q \propto P_{0}^{-1/5}$ and $\log q \propto M_{\rm cl}^{1/5}$, respectively.

Finally, we have predicted that there should exist a substantial fraction ($20-30$\%) of the total H$\alpha$ flux  coming from the faint, extended, old and evolved \HII\ regions with $\log q \le 7$cm s$^{-1}$, or $\log{\cal{U}} \le -3.5$. This emission has the appropriate flux, and should have the appropriate line ratios to be identified with the diffuse ionized gas (DIG) component of the ISM in galaxies.

 \begin{acknowledgements}
Dopita acknowledges the ANU and the Australian Research Council (ARC) for support of an ARC Federation Fellowship. Dopita, Sutherland \& Fishera acknowledge support through ARC Discovery project grant DP0208445. Kewley acknowledges a Hubble Fellowship. The work by van Breugel was performed under the auspices of the U.S. Department of Energy and Lawrence Livermore National Laboratory under contract No. W-7405-Eng-48.

\end{acknowledgements}

\end{document}